\documentclass[11pt,a4paper]{article}
\usepackage{jheppub}
\usepackage{bm,booktabs,mathrsfs}
\usepackage[section]{placeins}
\hypersetup{hidelinks}
\newcommand{\dd}{\mathrm{d}}
\newcommand{\ii}{\mathrm{i}}
\newcommand{\tr}{\operatorname{tr}}
\newcommand{\Tr}{\operatorname{Tr}}
\newcommand{\Slash}[1]{\not\!#1}
\newcommand{\cbar}{\bar c}
\newcommand{\R}{\mathcal R}
\newcommand{\Th}{\Theta}
\newcommand{\sym}{\operatorname{sym}}

\let\originaltableofcontents\tableofcontents
\renewcommand{\tableofcontents}{{\fontsize{9}{10}\selectfont\originaltableofcontents}}
\renewcommand{\afterTocRuleSpace}{\clearpage}

\title{Gravitational form factors of the baryon octet and flavor dependence in an NJL quark--diquark model}
\author[a]{Yiwei Qiu,}
\author[b]{Zhibo Liu}
\author[c]{and Congcong Zhang}
\affiliation[a]{Institute of Particle Physics and Key Laboratory of Quark and Lepton Physics (MOE),
Central China Normal University, Wuhan 430079, China}
\affiliation[b]{Kobayashi-Maskawa Institute for the Origin of Particles and the Universe,
Nagoya University, Nagoya 464-8602, Japan}
\affiliation[c]{Hunan Key Laboratory for Stellar and Interstellar Physics and School of Physics and Optoelectronics,
Xiangtan University, Hunan 411105, China}
\emailAdd{qiuyw@mails.ccnu.edu.cn}
\emailAdd{zhibo.liu.hep@gmail.com}
\emailAdd{zhangcongcong@xtu.edu.cn}
\abstract{We calculate the gravitational form factors $A$, $J$ and $D$
of the baryon octet in a Nambu--Jona-Lasinio quark--diquark model with
proper-time regularization. With the diquark poles used for the octet
spectrum, the $\Sigma$ has the smallest $|D(0)|$ and the largest $r_D$,
$0.675~\mathrm{fm}$ compared with $0.637~\mathrm{fm}$ for the nucleon.
Its $A$ radius is smaller than the nucleon's.
Along the strange-quark mass trajectory, $r_D^\Sigma-r_D^N$ is
negative at the lowest nonzero mass and positive at the other three.
The difference $r_A^\Sigma-r_A^N$ is negative at all four masses.
At a common spatial resolution,
the $\Sigma$ also has the weakest pressure and shear peaks.}

\begin{document}
\maketitle
\raggedbottom
\section{Introduction}
\label{sec:introduction}

The internal structure of a hadron is encoded in matrix elements of the
energy--momentum tensor (EMT) between hadron states. Lorentz covariance
and current conservation reduce these matrix elements to a small set of
gravitational form factors (GFFs), whose momentum dependence fixes the
distributions of momentum and angular momentum and, through the
gravitational form factor $D$, the internal stresses of the
hadron~\cite{Ji1997,Polyakov2003,Polyakov2018,Burkert2023}. The
spin-$1/2$ EMT decomposition was given in ref.~\cite{Pagels1966}.
Its connection to the pressure and shear distributions is reviewed
in ref.~\cite{Polyakov2018}.

For a conserved total EMT, Poincar\'e invariance gives
$A(0)=1$, $J(0)=1/2$ and $B(0)=0$~\cite{Brodsky2001,Lowdon2017,LorceLowdon2020,CaoBN2026},
while $D$ and the form-factor slopes describe the internal dynamics.
Improvement terms preserve the conserved charges and modify the local
EMT~\cite{Callan1970}. Curvature couplings contribute to its metric
derivative through the transverse part of the gravitational
vertex~\cite{Hudson2017,Alharazin2020}.

GFFs are accessible through hard exclusive reactions because they
appear as moments of generalized parton
distributions~\cite{JiDVCS1997,Radyushkin1996}, whose theory is summarized
in~\cite{Diehl2003,Belitsky2005}. Analyses of deeply virtual Compton
scattering data have extracted the proton's quark stress
distribution~\cite{Burkert2018}. Lattice QCD has determined quark and
gluon contributions to proton momentum and stress
distributions~\cite{Shanahan2019,Hackett2024}, and gluon GFFs have been
compared across hadrons of different
spin~\cite{ShanahanGluon2019,Pefkou2022}. Symmetry-preserving continuum
equations provide a unified nucleon calculation~\cite{Yao2025}, while
dispersive methods relate the stress response to low-energy hadronic
amplitudes~\cite{CaoGuo2025,CaoGuoReview2026}. Long-distance pion and
$\Delta$ dynamics are expected to modify these
responses~\cite{Wang2024,Alharazin2024}.
Generative diffusion models also provide non-parametric reconstructions
of the proton GFFs from sparse lattice
data~\cite{AlharazinPanteleeva2026}.

The baryon octet allows a comparison of strangeness and diquark
correlations in hadron structure. The
$\Lambda$ and $\Sigma$ have the same strange-valence content and different
spin--flavor wave functions, so comparing them, together with the nucleon
and $\Xi$, probes the spin and flavor dependence of their internal
correlations. The chiral quark--soliton model has provided EMT form factors of the nucleon
and their flavor
dependence~\cite{Goeke2007,Won2022,Won2024}. Light-cone sum rules give
hyperon and nucleon GFFs from quark
operators~\cite{Ozdem2020,Azizi2020}. Holographic QCD predicts a
decrease of the momentum radii with increasing strangeness~\cite{Liu2026}.
The normalized $D$ radius is determined by $D(0)$ and its slope.

The present calculation uses a proper-time-regulated
Nambu--Jona-Lasinio (NJL) framework, in which dynamical chiral symmetry
breaking generates the constituent masses~\cite{Nambu1961,Vogl1991}, with
the model applied to mesons and baryons
in~\cite{Klevansky1992,Hatsuda1994}. Baryons appear as solutions of a
relativistic Faddeev equation for quarks and
diquarks~\cite{Ishii1995}, and the evidence for diquark correlations in
hadron structure is reviewed in~\cite{Barabanov2021}. The octet is
described with a common parameter
set~\cite{Carrillo2016,CarrilloThesis}. Earlier NJL studies obtain meson
GFFs~\cite{Freese2019,FreeseErratum2022} and proton GFFs from dressed
generalized parton distributions~\cite{Freese2020}. Continuum and
contact-interaction calculations emphasize dynamics beyond the quark
gravitational Ward--Takahashi identity~\cite{Freese2019,Xu2024,Sultan2024}.
Mesons containing light and strange quarks have been studied in holographic QCD and proper-time NJL
models~\cite{LiuKaon2025,LiuAbuki2026}. The interaction contribution to
the EMT is explicit in nonperturbative scalar--Yukawa
studies~\cite{CaoLiVary2023,CaoLiVary2024}. Diquark correlations and the
pion cloud are known to reshape nucleon electromagnetic form
factors~\cite{Cloet2014}. Here we obtain the EMT by varying the
action with respect to the background geometry. We include the
interaction terms, the transitions between scalar and axial-vector
diquarks, and the variation of the spin connection.

The gravitational form factor $D$ determines the Breit-frame pressure
and shear distributions~\cite{Lorce2019,Jaffe2021}.
Their relation to light-front force distributions can be expressed
through an Abel transform~\cite{Panteleeva2021}.
Definitions based on localized wave packets have been developed for
electromagnetic and gravitational
densities~\cite{Panteleeva2022EM,Panteleeva2023GFF}.
Light-front densities have also been studied
in refs.~\cite{FreeseMiller2021,FreeseMiller2022}, and mechanical
stability conditions in ref.~\cite{GegeliaPolyakov2021}.
In scale-invariant Skyrme models the field
associated with the gluonic scale anomaly generates an extended
attractive stress and a substantial $D$
term~\cite{FujiiKawaguchi2025,TanakaKawaguchi2025}. The $D$ form factor has
been related to the chirally invariant part of the nucleon
mass~\cite{Kawaguchi2026}. The trace--traceless decomposition of the
QCD EMT resolves the corresponding components~\cite{FujiiTanaka2025}.

We calculate $A$, $J$ and $D$ for the baryon octet and compare the
radii obtained from $A$ and $D$. To follow the origin of the splittings,
we vary the strange current-quark mass from the flavor-symmetric point
to its physical value, solving the diquark and baryon equations along
the trajectory. We also examine how the radii change with the diquark
inputs and the transverse part of the gravitational current.

Section~\ref{sec:model} defines the bound states and the total EMT.
Section~\ref{sec:results} presents the octet form factors and the flavor
dependence. Section~\ref{sec:flavor-breaking} follows its development
under flavor breaking, and section~\ref{sec:mechanics} gives
the pressure and shear distributions at a common spatial resolution.


\section{Quark--diquark framework and the energy--momentum tensor}
\label{sec:model}

\subsection{Interaction and proper-time regularization}

We take the isospin limit, $M_u=M_d=M_l$, and use
$g^{\mu\nu}=\operatorname{diag}(1,-1,-1,-1)$.
The NJL interaction implements dynamical mass generation and
spontaneous chiral symmetry breaking through local quark
interactions~\cite{Nambu1961,Vogl1991}.
We use the three-flavor NJL Lagrangian of Ref.~\cite{Carrillo2016},
\begin{align}
\mathcal L={}&\bar\psi(\ii\Slash\partial-\hat m)\psi
+\frac{G_\pi}{2}\sum_{j=0}^{8}
\left[(\bar\psi\lambda_j\psi)^2
-(\bar\psi\gamma_5\lambda_j\psi)^2\right]\nonumber\\
&-\frac{G_\rho}{2}\sum_{j=0}^{8}
\left[(\bar\psi\gamma^\mu\lambda_j\psi)^2
+(\bar\psi\gamma^\mu\gamma_5\lambda_j\psi)^2\right].
\label{eq:lag}
\end{align}
Here $\hat m=\operatorname{diag}(m_l,m_l,m_s)$,
$\lambda_0=\sqrt{2/3}\,I$, and
$\tr_f\lambda_j\lambda_k=2\delta_{jk}$.
The scalar and axial-vector diquark couplings, $G_s$ and $G_a$,
are independent parameters.

For a Euclidean denominator, the regulator is defined by
\begin{equation}
\frac{1}{(\ell_E^2+\Omega)^n}
\longrightarrow
\frac{1}{\Gamma(n)}
\int_a^b\dd\tau\,\tau^{n-1}e^{-\tau(\ell_E^2+\Omega)},
\label{eq:pt}
\end{equation}
where $a=\Lambda_{\rm UV}^{-2}$ and $b=\Lambda_{\rm IR}^{-2}$.
We use
\begin{align}
I_1(\Omega)&=\frac{1}{16\pi^2}\int_a^b
\frac{\dd\tau}{\tau^2}e^{-\tau\Omega},\nonumber\\
I_2(\Omega)&=\frac{1}{16\pi^2}\int_a^b
\frac{\dd\tau}{\tau}e^{-\tau\Omega},\nonumber\\
I_0(\Omega)&=\frac{1}{16\pi^2}\int_a^b
\dd\tau\,e^{-\tau\Omega}.
\label{eq:integrals}
\end{align}
Here $I_1'=-I_2$ and $I_2'=-I_0$.
We use the same infrared and ultraviolet cutoffs for the
propagators and gravitational vertices.

The Hartree gap equation is
\begin{equation}
M_f=m_f+16N_cG_\pi M_f I_1(M_f^2),\qquad N_c=3.
\label{eq:gap}
\end{equation}
We fix the constituent masses
at $M_l=0.400$ and
$M_s=0.560~\mathrm{GeV}$ and use
\begin{align}
\Lambda_{\rm UV}&=0.645~\mathrm{GeV},&
\Lambda_{\rm IR}&=0.240~\mathrm{GeV},\nonumber\\
G_\pi&=19.0~\mathrm{GeV}^{-2},&
G_\rho&=11.0~\mathrm{GeV}^{-2},\nonumber\\
G_s&=5.8~\mathrm{GeV}^{-2},&
G_a&=4.9~\mathrm{GeV}^{-2}.
\label{eq:parameters}
\end{align}
With these constituent masses and regulator, Eq.~\eqref{eq:gap}
fixes the current masses. We use the physical-pion-mass
calibration~\cite{Carrillo2016} for the baryon spectrum and the
lattice comparison below.

\subsection{Diquark two-point functions}

The retained diquark channels consist of flavor-antisymmetric
scalar and flavor-symmetric axial-vector correlations.
Their normalized flavor states are
$[fg]=(|fg\rangle-|gf\rangle)/\sqrt2$,
$\{fg\}=(|fg\rangle+|gf\rangle)/\sqrt2$, and $\{ff\}=|ff\rangle$.
The color matrices are
$\beta_A=\sqrt{3/2}\lambda_A^{(c)}$, $A=2,5,7$.
These conventions fix the color and flavor factors in the diquark
vertices and in the Faddeev kernel.

For quark masses $M_1,M_2$ and diquark virtuality $s$, define
\begin{equation}
\Omega_{12}=xM_1^2+(1-x)M_2^2-x(1-x)s.
\end{equation}
The scalar polarization and the transverse and longitudinal
axial-vector polarizations are
\begin{align}
\Pi_s(s)={}&-12\left\{I_1(M_1^2)+I_1(M_2^2)\right.\nonumber\\
&\left.+[s-(M_1-M_2)^2]\int_0^1\dd x\,I_2(\Omega_{12})\right\},
\label{eq:pis}\\
\Pi_T(s)={}&24\int_0^1\dd x\,
\left[xM_1^2+(1-x)M_2^2-M_1M_2\right.\nonumber\\
&\left.\hspace{34mm}-2x(1-x)s\right]I_2(\Omega_{12}),
\label{eq:pit}\\
\Pi_L(s)={}&24\int_0^1\dd x\,
[xM_1^2+(1-x)M_2^2-M_1M_2]I_2(\Omega_{12}).
\label{eq:pil}
\end{align}
The longitudinal polarization vanishes in the equal-mass case,
whereas it generally contributes for unequal quark masses.

The resummed scalar scattering matrix and pole condition are
\begin{align}
\tau_s(s)&=\frac{4\ii G_s}{1+2G_s\Pi_s(s)},\nonumber\\
1+2G_s\Pi_s(m_d^2)&=0,\qquad
Z_d=-\frac{2}{\Pi_s'(m_d^2)}.
\label{eq:rpa}
\end{align}
For the axial-vector transverse pole, replace $G_s,\Pi_s$ by
$G_a,\Pi_T$. We solve these pole equations at each strange-quark
mass in section~\ref{sec:flavor-breaking}.

For the main form-factor calculation we use the diquark pole masses
and residues of ref.~\cite{Carrillo2016}. Each propagator contains
a momentum-independent term and a pole term. After removing an
overall phase, the propagators are
\begin{align}
\Delta_s(k)&=c_s+\frac{Z_s}{m_s^2-k^2},\qquad c_s=4G_s,\nonumber\\
\Delta_a^{\alpha\beta}(k)
&=c_ag^{\alpha\beta}
+\frac{Z_a}{m_a^2-k^2}
\left(g^{\alpha\beta}-\frac{k^\alpha k^\beta}{m_a^2}\right),
\label{eq:diquark-propagators}\\
c_a&=4G_a.\nonumber
\end{align}
The scalar pole inputs are $m_d=(0.768,0.903)~\mathrm{GeV}$ and $Z_d=(11.10,12.00)$ for $[ll]$ and $[ls]$. The axial inputs are $m_d=(0.929,1.040,1.150)~\mathrm{GeV}$ and $Z_d=(6.73,7.54,8.36)$ for $\{ll\}$, $\{ls\}$ and $\{ss\}$.
In the diquark propagators, $m_s$ and $m_a$ denote the scalar and
axial-vector pole masses. We identify the strange current-quark
mass through $\hat m$ or label it explicitly in the text.
We retain the complete axial-vector pole numerator in
Eq.~\eqref{eq:diquark-propagators}. 

\subsection{Faddeev equation and baryon masses}

The quark--diquark description of baryons has been developed
through relativistic Faddeev equations~\cite{Ishii1995}.
Evidence for these correlations in hadron structure is reviewed
in Ref.~\cite{Barabanov2021}.
In the static approximation of ref.~\cite{Carrillo2016}, the
propagator of the exchanged quark is replaced by $-1/M_f$, where
$M_f$ is its constituent mass.
Writing the exchange kernel as $\mathcal Z_{\rm ex}$, the
homogeneous equation is
\begin{equation}
\Gamma_0=\mathcal Z_{\rm ex}\Pi_{qd}\Gamma_0.
\label{eq:faddeev}
\end{equation}
For the $N,\Sigma,\Xi$ families the vertex has five covariants,
\begin{equation}
\Gamma_0(p)u=
\begin{pmatrix}
\alpha_1 I\\
\alpha_2p^\alpha\gamma_5/M_B+\alpha_3\gamma^\alpha\gamma_5\\
\alpha_4p^\alpha\gamma_5/M_B+\alpha_5\gamma^\alpha\gamma_5
\end{pmatrix}u .
\label{eq:vertex}
\end{equation}
The $\Lambda$ has two scalar amplitudes and one axial-vector
pair. We use the flavor assignments of ref.~\cite{Carrillo2016}
for the exchange kernel.

The scalar quark--diquark loop includes the momentum-independent term,
\begin{align}
\Pi_{qd,s}(p)={}&Z_s\int_0^1\dd x\,
[m+(1-x)\Slash p]I_2(\Omega)\nonumber\\
&+4G_s m I_1(m^2),\nonumber\\
\Omega={}&xm^2+(1-x)m_s^2-x(1-x)p^2.
\label{eq:qdscalar}
\end{align}
Here $m$ is the spectator constituent-quark mass.
For the axial-vector tensor reduction, we use
\begin{equation}
L_{2,\mathrm{alg}}(\Omega)=I_1(\Omega)-\Omega I_2(\Omega).
\label{eq:lalg}
\end{equation}
Direct Gaussian integration of the corresponding rank-two numerator
gives $L_{2,\mathrm{G}}=2I_1$. The two results differ by a term
from the proper-time integration limits,
\begin{align}
\delta L(\Omega)&=L_{2,\mathrm{alg}}-L_{2,\mathrm{G}}\nonumber\\
&=\frac{-e^{-a\Omega}/a+e^{-b\Omega}/b}{16\pi^2}.
\label{eq:deltaL}
\end{align}
For a spectator mass $m$ and an axial-vector diquark mass $d$,
let $N=m+(1-x)\Slash p$ and
$\Omega=xm^2+(1-x)d^2-x(1-x)p^2$.
The axial-vector quark--diquark loop is
\begin{align}
\Pi_a^{ab}(p)={}&4G_amI_1(m^2)g^{ab}\nonumber\\
&+Z_a\int_0^1\dd x
\left\{N\left[g^{ab}I_2-\frac{x^2p^ap^b}{d^2}I_2
+\frac{g^{ab}}{4d^2}L_2\right]\right.\nonumber\\
&\left.\hspace{8mm}-\frac{xL_2}{4d^2}
(p^a\gamma^b+p^b\gamma^a)\right\}.
\label{eq:axialbubble}
\end{align}
Here $L_2=L_{2,\mathrm{alg}}$. The term from the integration limits
that enters the metric variation of this loop is
\begin{align}
\Sigma^{ab}(p)
={}&\frac{Z_a}{4d^2}\int_0^1\dd x\,\delta L(\Omega)
\left\{g^{ab}[m+(1-x)\Slash p]\right.\nonumber\\
&\left.\hspace{20mm}-x(p^a\gamma^b+p^b\gamma^a)\right\}.
\label{eq:endpointkernel}
\end{align}

The baryon mass is determined by an eigenvalue of
$\mathcal Z_{\rm ex}\Pi_{qd}$ crossing one.
The coefficients satisfy $\alpha^T\mathsf G\alpha=1$, where each
axial-vector block of the normalization matrix $\mathsf G$ is
\begin{equation}
\mathsf G_{\rm ax}=\begin{pmatrix}1&-1\\-1&4\end{pmatrix}.
\end{equation}
The pole normalization used in the EMT matrix element is
\begin{equation}
Z_B^{-1}=2M_B\bar\Gamma_0
\left.\frac{\partial\Pi_{qd}}{\partial p^2}\right|_{p^2=M_B^2}
\Gamma_0 .
\label{eq:zb}
\end{equation}
The derivative acts on $\Pi_{qd}$. We continue the left vertex
covariantly to complex momentum. The pole residue $Z_B$ sets the
current normalization.


\subsection{Energy--momentum tensor}
\label{sec:sources}

We obtain the total EMT by varying the regulated quark action and
the effective diquark operators. The canonical quark EMT
is~\cite{Freese2019}
\begin{equation}
 T_C^{\mu\nu}=\bar\psi\ii\gamma^\mu
 \overleftrightarrow\partial^{\,\nu}\psi
 -g^{\mu\nu}\left[\bar\psi(\ii\overleftrightarrow{\Slash\partial}
 -\hat m)\psi+\mathcal L_I\right],
 \label{eq:canonical}
\end{equation}
where $\overleftrightarrow\partial=(\overrightarrow\partial-
\overleftarrow\partial)/2$. Interaction contractions and the gap
equation replace the current mass by the constituent mass in the
quark vertex. The metric variation also acts on the torsion-free
spin connection,
\begin{equation}
 W_H[e]=W[e,\omega_{\rm LC}(e)],\qquad
 \delta W_H=\left.\delta W\right|_\omega+
 \frac{\delta W}{\delta\omega}\delta\omega_{\rm LC}.
 \label{eq:hilbert}
\end{equation}
We write $\widehat D=g^{1/4}D_{\rm geom}g^{-1/4}$ and define the
quark-loop contribution by
\begin{equation}
 W_{\rm even}=\frac{N_c}{2}\int_a^b\frac{\dd\tau}{\tau}
 \Tr e^{-\tau H},\qquad H=\widehat D^\dagger\widehat D.
 \label{eq:det}
\end{equation}
The mixed scalar--axial response also retains the phase
\begin{align}
 W_{\rm odd}={}&-\frac{N_c}{2}\int_0^1\dd\zeta\int_a^b\dd\tau\,
 \Tr\!\left[e^{-\tau H_\zeta}
 (\widehat D_\zeta^\dagger\mathcal P-
 \mathcal P^\dagger\widehat D_\zeta)\right],\nonumber\\
 \widehat D_\zeta={}&\widehat D_{\rm base}+\zeta\mathcal P,
 \qquad H_\zeta=\widehat D_\zeta^\dagger\widehat D_\zeta .
 \label{eq:odd-phase}
\end{align}
Here $\mathcal P$ is the $\gamma_5$ scalar-diquark insertion in the
charge-conjugated representation. The scalar diquark has parity
$0^+$. Both functionals use the same proper-time interval. Their
derivatives preserve the order of the operator products and include
the contact terms and the tree-level contribution of the auxiliary fields.
For the scalar field
of flavor $f$, define
$\Omega=M_f^2+x(1-x)Q^2$. The regulated quadratic kernel and
response to the metric variation are
\begin{align}
K_f(Q^2)={}&\frac{1}{4G_\pi}-4N_cI_1(M_f^2)\nonumber\\
&+2N_c(4M_f^2+Q^2)\int_0^1\dd x\,I_2(\Omega),\nonumber\\
L_f(Q^2)={}&8N_cM_f\int_0^1\dd x\,
(x-\tfrac12)^2 I_2(\Omega),\nonumber\\
\chi_f(Q^2)={}&-\frac{L_f(Q^2)}{K_f(Q^2)} .
\label{eq:chi}
\end{align}
The transverse scalar contribution to the quark vertex is
$\Th^{\mu\nu}\chi_f$.

The diquark vertex for Eq.~\eqref{eq:diquark-propagators} includes
the transverse contribution from the quark loop.
For $s=p^2$, $\delta=s-m_d^2$, the scalar inverse is
\begin{equation}
 F_C(s)=\frac{Z_s\delta}{Z_s-c_s\delta},\qquad
 F_C\Delta_C=-Z_s,
 \label{eq:fc}
\end{equation}
and the vector inverse is
\begin{align}
 F_C^{ab}&=f_C(s)g^{ab}+h_C(s)p^ap^b,\nonumber\\
 f_C(s)&=-\frac{Z_a\delta}{Z_a-c_a\delta},\qquad
 h_C(s)=\frac{Z_a^2}{(Z_a+c_am_a^2)(Z_a-c_a\delta)}.
 \label{eq:axialinverse}
\end{align}
It satisfies $F_C^{ac}\Delta_{C,c}^{\phantom{C,c}b}=Z_ag^{ab}$,
including the longitudinal part. In the same pole normalization,
the inverse scattering matrices follow from the polarizations in
Eqs.~\eqref{eq:pis}--\eqref{eq:pil}:
\begin{align}
 F_{{\rm mic},s}&=-\frac{Z_s}{4G_s}(1+2G_s\Pi_s),\nonumber\\
 F_{{\rm mic},a}^{ab}&=\frac{Z_a}{4G_a}
 [(1+2G_a\Pi_T)P_T^{ab}+(1+2G_a\Pi_L)P_L^{ab}],
 \label{eq:mic-inverse}
\end{align}
where $P_L^{ab}=p^ap^b/s$ and $P_T^{ab}=g^{ab}-P_L^{ab}$.
For a scalar inverse $F$, define
\begin{align}
 V[F]^{\mu\nu}={}&(2K^\mu K^\nu-\Th^{\mu\nu}/2)
 F^{[1]}(s_f,s_i)-\tfrac12g^{\mu\nu}[F(s_f)+F(s_i)],\nonumber\\
 F^{[1]}(s_f,s_i)={}&\frac{F(s_f)-F(s_i)}{s_f-s_i},\qquad
 \Th^{\mu\nu}=q^\mu q^\nu-g^{\mu\nu}q^2,
 \label{eq:minimum}
\end{align}
with $K=(p_f+p_i)/2$ and the derivative limit at coincident
arguments. Let $\Gamma_{\rm mic}$ be the vertex obtained from the
metric variation of the quark loop. The diquark vertex is
\begin{equation}
 \Gamma_C=V[F_C]+R,\qquad
 R=\Gamma_{\rm mic}-V[F_{\rm mic}].
 \label{eq:matching}
\end{equation}
For the axial-vector diquark, the divergence and spin connection
give
\begin{align}
 V^{\mu\nu;ab}[f,h]={}&g^{ab}V[f]^{\mu\nu}
 +p_f^ap_i^bV[h]^{\mu\nu}\nonumber\\
 &+p_f^\nu g^{\mu a}p_i^b h(s_i)
 +p_i^\nu p_f^ag^{\mu b}h(s_f).
 \label{eq:vectormin}
\end{align}
In Euclidean space, let $H_s=d^2+L_s$ and $H_v=d^2+L_v$, with
$L_s$ and $L_v$ the covariant scalar and vector Laplacians.
Their flat-space values are $p_E^2$ and $p_E^2I$. For the
axial-vector diquark, define $J_s=Z+cH_s$, $J_v=Z+cH_v$ and
$\kappa=Z^2/(Z+cd^2)$, where $d$ is its mass, $Z$ its pole
residue, and $c$ its momentum-independent coefficient. With $C$
the covariant divergence, the main calculation uses
$F_a^{(s)}$ in
\begin{align}
 F_a^{(s)}&=ZH_vJ_v^{-1}-\kappa C^\dagger J_s^{-1}C,\nonumber\\
 F_a^{(v)}&=ZH_vJ_v^{-1}-\frac{\kappa}{2}\{C^\dagger C,J_v^{-1}\}.
 \label{eq:axial-orderings}
\end{align}
The second operator gives the same flat-space propagator and
a different metric variation. Curvature terms in chiral effective
theory illustrate this freedom in the coupling to
gravity~\cite{Hudson2017,Alharazin2020,LiCaoGuo2026}.

The scalar contribution from $R_s$ has the loop weight
\begin{equation}
 W_R=\frac{c_s^2}{Z_s}\R_1(H_q)+c_s\R_2(H_q,H_f)
 +c_s\R_2(H_q,H_i)+Z_s\R_3(H_q,H_f,H_i),
 \label{eq:wr}
\end{equation}
where $H_{f,i}=m_s^2-k_{f,i}^2$ and $\R_n$ is defined in
Eq.~\eqref{eq:rn}.
We obtain the baryon EMT from the metric variation of the
inverse Faddeev kernel. This applies the gauging of equations
method~\cite{Kvinikhidze1999} to an external gravitational field,
\begin{equation}
 \mathscr K_B=\mathcal Z_{\rm ex}^{-1}-\Pi_{qd},\qquad
 \Gamma_B^{\mu\nu}=Z_B\bar\Gamma_0
 \mathcal D_H^{\mu\nu}\mathscr F_{\rm sel}\Gamma_0,
 \qquad \mathscr F_{\rm sel}|_0=\mathscr K_B .
 \label{eq:fullkernel}
\end{equation}
For the scalar term we remove the overall loop phase and use
\begin{equation}
 W_s[X]=\left[\int\frac{\dd^4r}{(2\pi)^4}
 \frac{\Slash r+M_q}{H_q}
 \frac{\Delta_{C,s}(k_f)X\Delta_{C,s}(k_i)}{Z_s}\right]_{[a,b]},
 \qquad H_q=M_q^2-r^2.
 \label{eq:scalar-sewing}
\end{equation}
In the presence of the metric source, we use the kernel
\begin{align}
 \mathscr F_{\rm sel}={}&\mathcal Z_{\rm ex}^{-1}
 -\Pi_{\rm ref}-\Pi_{\rm end}
 -W_s[\mathcal C_s]-W_a[\mathcal C_a]\nonumber\\
 &-W_{\rm diff}[\mathcal E]-W_{sa}[\mathcal T].
 \label{eq:selected-parent}
\end{align}
The operators $\mathcal C_d$ vanish in flat space and have first
variations $\mathcal D_H\mathcal C_d|_0=-R_d$ for $d=s,a$.
Thus the variation of $-W_d[\mathcal C_d]$ adds $W_d[R_d]$ to
the current. The term $\Pi_{\rm end}$ follows from
Eq.~\eqref{eq:endpointkernel}. The loop $\Pi_{\rm ref}$ uses the
Euclidean axial-vector propagator
\begin{equation}
 G_{\rm ref}=cI+\frac{Z}{2}\{I+C_{1/2}^{\dagger}C_{1/2}/d^2,H_v^{-1}\}.
\end{equation}
In Euclidean space $C_{1/2}=C$ at zero source.
For a variation $u_a{}^b=\delta e_a{}^b$ carrying momentum $q$,
the first variation of $C_{1/2}$ is
\begin{equation}
 (\delta C_{1/2})_{fi,a}
 =\frac{1}{2}\left[u_a{}^b(p_f+p_i)_b+\sum_c\Omega_{c,ca}\right],
\end{equation}
where $\Omega$ is the vector spin connection after removing the
Fourier phase. The propagator $G_a=Z(F_a^{(s)})^{-1}$ enters
Eq.~\eqref{eq:selected-parent}. With $J_v=Z+cH_v$, the operator
$\mathcal E=H_v(J_v/c)(G_a-G_{\rm ref})(J_v/c)H_v$ accounts for
the difference in the metric variations of the two propagators.
Their difference vanishes in flat space. The loop $W_{\rm diff}$
multiplies $\delta\mathcal E$ by
$\R_5(H_q,H_f,H_i,H_f+Z/c,H_i+Z/c)$, with
$H_{f,i}=d^2-k_{f,i}^2$ for the axial-vector diquark.
The operator $\mathcal T$
collects the two transitions between scalar and axial-vector diquarks.
The two quark insertions in each transition have opposite flavor
weights. Their loop contributions cancel when the two masses are
equal. We include the pole residues in the external diquark
propagators and use unit residues inside the transition loop.

The variation that gives the symmetric EMT is
\begin{equation}
 \mathcal D_H=-\sym\left[\delta_e+
 \sum_f(\delta_e\sigma_f)\delta_{\sigma_f}
 +(\delta_e\omega_{\rm LC})\delta_\omega\right],\qquad
 \delta_e\bm\sigma=-K_{\sigma\sigma}^{-1}K_{\sigma e}.
 \label{eq:parent-source-chain}
\end{equation}
The diquark pole masses, residues and momentum-independent coefficients
are taken as constants in this derivative. The induced scalar field
contributes to the quark propagators, including the spectator and the
quark exchange kernel. We use the same proper-time limits for the
combined denominator of each loop,
\begin{equation}
 \R_n(H_1,\ldots,H_n)=\int_{\Delta_{n-1}}\dd\bm x
 \int_a^b\dd\tau\,\tau^{n-1}e^{-\tau\sum_jx_jH_j},
 \qquad x_j\geq0,\quad\sum_jx_j=1.
 \label{eq:rn}
\end{equation}
We sum the metric variations of the quark and diquark propagators,
the interaction kernel and the spin connection before projecting
onto the form factors.

\subsection{Form factors and radii}

For $P=(p_f+p_i)/2$, $q=p_f-p_i$ and $t=q^2=-Q^2$, we use
\begin{align}
 \langle p_f|T_H^{\mu\nu}|p_i\rangle
 =\bar u_f\Big[&\tfrac12 A(t)(\gamma^\mu P^\nu+\gamma^\nu P^\mu)
 +\frac{B(t)}{4M_B}
 (\ii\sigma^{\mu\rho}q_\rho P^\nu+
  \ii\sigma^{\nu\rho}q_\rho P^\mu)\nonumber\\
 &+\frac{D(t)}{4M_B}\Th^{\mu\nu}
 +M_B\cbar(t)g^{\mu\nu}\Big]u_i,
 \label{eq:gff}
\end{align}
with $J=(A+B)/2$. For QCD EMT operators, the corresponding
form factors are related to moments of generalized parton
distributions~\cite{Belitsky2005}.
The normalized conserved total EMT obeys
$A(0)=1$, $J(0)=1/2$, $B(0)=0$ and
$\cbar(t)=0$~\cite{Ji1997,Brodsky2001,Lowdon2017,LorceLowdon2020}. At finite transfer,
$B=2J-A$ contains independent dynamical information, also examined
in light-front holographic QCD~\cite{CaoBN2026}.
Constituent contributions to the EMT are studied separately in
scalar--Yukawa models~\cite{CaoLiVary2024}.

For $F=A,D$, we define the normalized slope radius
\begin{equation}
 r_F^2=-\frac{6(\hbar c)^2}{F(0)}
 \left.\frac{\dd F(Q^2)}{\dd Q^2}\right|_{0}.
 \label{eq:slope-radii}
\end{equation}
The $D$ slope radius measures the normalized variation at $Q^2=0$.
The mechanical radius is
\begin{equation}
 r_{\rm mech}^2=\frac{6D(0)}{\int_0^\infty\dd Q^2\,D(Q^2)}
 \label{eq:mechanical-vs-slope-radius}
\end{equation}
in natural units~\cite{Polyakov2018}. The energy-density radius
follows from $A+Q^2(D-B)/(4M_B^2)$.


\section{Gravitational form factors and octet flavor structure}
\label{sec:results}

\subsection{Form factors of the baryon octet}

We calculate the baryon masses, $D(0)$ and slope radii in
Table~\ref{tab:octet-summary} using the diquark pole masses and
residues of ref.~\cite{Carrillo2016} as inputs.
\begin{table}[tb]
\centering
\small
\caption{Baryon masses $M$ (GeV), $D(0)$ and slope radii $r_A$ and $r_D$ (fm), all calculated in this work.}
\label{tab:octet-summary}
\begin{tabular}{lrrrr}
\toprule
Baryon & $M$ & $D(0)$ & $r_A$ & $r_D$\\
\midrule
$N$ & 0.9446 & -1.392 & 0.423 & 0.637\\
$\Lambda$ & 1.1259 & -1.392 & 0.401 & 0.624\\
$\Sigma$ & 1.1779 & -1.091 & 0.416 & 0.675\\
$\Xi$ & 1.3205 & -1.127 & 0.394 & 0.615\\
\bottomrule
\end{tabular}
\end{table}

\begin{figure}[tb]
\centering
\includegraphics[width=.76\textwidth]{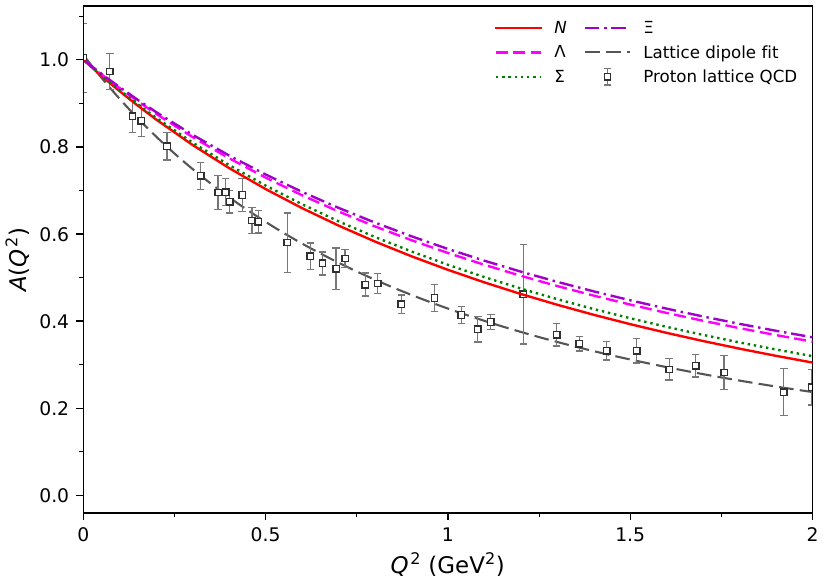}
\caption{Gravitational form factor $A$ for $N$ (red solid),
$\Lambda$ (magenta dashed), $\Sigma$ (green dotted) and $\Xi$
(purple dash-dotted). Symbols are the proton total-EMT lattice
results of ref.~\cite{Hackett2024}.}
\label{fig:gff}
\end{figure}
\begin{figure}[tb]
\centering
\includegraphics[width=.76\textwidth]{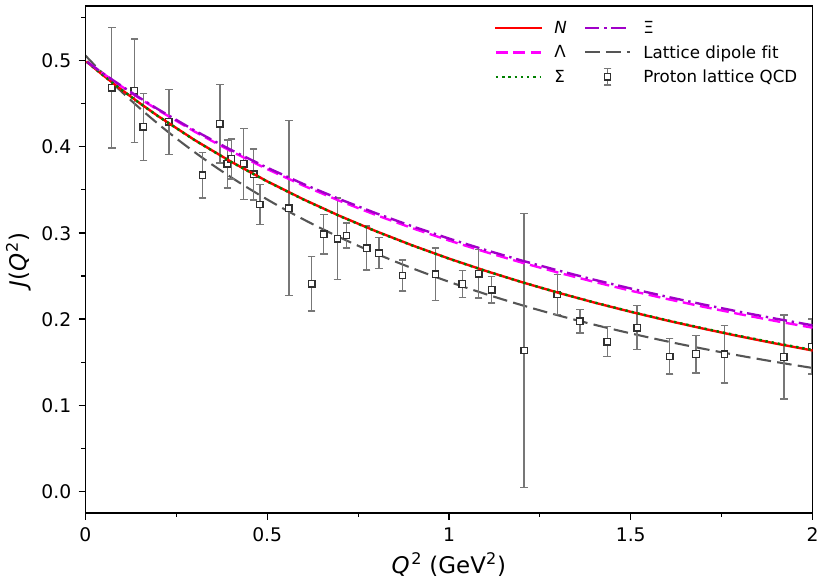}
\caption{Angular-momentum form factor $J(Q^2)$, with baryon and
lattice conventions as in Fig.~\ref{fig:gff}.}
\label{fig:gff-J}
\end{figure}
\begin{figure}[tb]
\centering
\includegraphics[width=.76\textwidth]{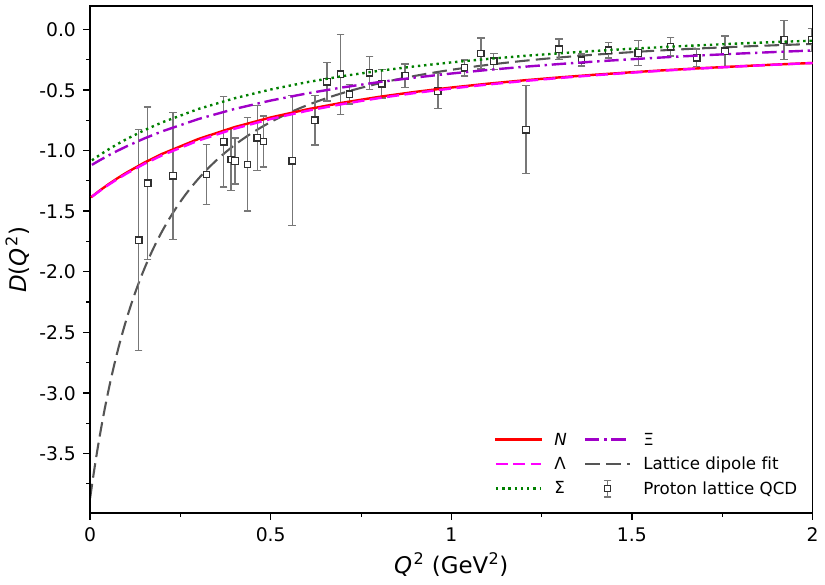}
\caption{Gravitational form factor $D$, with the conventions of
Fig.~\ref{fig:gff}.}
\label{fig:gff-D}
\end{figure}

We evaluate the complete current in Breit kinematics over
$0.005\leq Q^2\leq2~\mathrm{GeV}^2$. Figures~\ref{fig:gff}--\ref{fig:gff-D}
show the results for the four baryons. Both $A$ and $J$ decrease
with $Q^2$ for each baryon. The gravitational form factor $D$ is
negative throughout this interval, with the smallest magnitude
in the $\Sigma$ channel.


\subsection{Ordering of the slope radii}
\label{sec:slope-radii}
\label{sec:radius-ordering}

With the diquark poles from the spectrum calculation, we obtain
\begin{equation}
 r_A^N>r_A^\Sigma>r_A^\Lambda>r_A^\Xi,
 \qquad r_D^\Sigma>r_D^N>r_D^\Lambda>r_D^\Xi.
 \label{eq:radius-ordering}
\end{equation}
The $\Sigma$ has $r_D=0.675~\mathrm{fm}$, which is
$0.038~\mathrm{fm}$ larger than the nucleon value. Its $D(0)=-1.091$
is closer to zero than the nucleon's $-1.392$
(Table~\ref{tab:octet-summary}).

Separate $\pm5\%$ variations of $G_s$ and $G_a$, with the diquark
and baryon equations solved again, preserve both radius orderings.
Using
$F_a^{(v)}$ in Eq.~\eqref{eq:axial-orderings} reverses the
$N$--$\Lambda$ order for $r_D$ and preserves the $r_A$ ordering.
Changing the input diquark pole masses can change both orderings.
Across these parameter and vertex choices, the $\Sigma$ has
larger $r_A$ and $r_D$ than the $\Xi$.

\subsection{Relation to proton and hyperon calculations}
\label{sec:lattice-discussion}

Figures~\ref{fig:gff}--\ref{fig:gff-D} show the proton lattice results
at $m_\pi\simeq170~\mathrm{MeV}$~\cite{Hackett2024}. The NJL calculation
uses parameters calibrated at the physical pion mass. Applying
Eq.~\eqref{eq:slope-radii} to the total form factors obtained from the
lattice dipole fits gives $r_A^p\simeq0.503~\mathrm{fm}$ and
$r_D^p\simeq1.145~\mathrm{fm}$. The corresponding NJL radii are
$0.423~\mathrm{fm}$ and $0.637~\mathrm{fm}$. The lattice fits give
$D_N(0)=-3.87(97)$ for the dipole and $-3.35(58)$ for the
$z$-expansion, compared with $-1.392$ in the model.

The continuation to $t=0$ is sensitive to the large uncertainties
at low $Q^2$. Alharazin and Panteleeva reconstruct the same lattice
form factors with a generative diffusion
model~\cite{AlharazinPanteleeva2026}. At the lattice pion mass,
the 68\% band at $t=0$ includes our value.

Continuum and dispersive determinations also give more negative
values of the proton gravitational form factor $D(0)$, approximately $-3.11$ and $-3.38$
respectively~\cite{Yao2025,CaoGuo2025}.
Chiral matching of the diffusion reconstruction gives
$D_N(0)=-4.3\pm0.8$ at the physical pion
mass~\cite{AlharazinPanteleeva2026}. Long-distance pion and
$\Delta$ dynamics can modify this response~\cite{Wang2024,Alharazin2020,Alharazin2024}.
In scale-invariant Skyrme models, the scalar
field associated with the gluonic scale anomaly contributes an
extended attractive stress~\cite{FujiiKawaguchi2025,TanakaKawaguchi2025}.
Under light-scalar dominance, $D$ has also been related to the
chirally invariant nucleon mass~\cite{Kawaguchi2026}.

The chiral quark-soliton approach gives nucleon
EMT form factors~\cite{Goeke2007} and the octet values
$(D_N,D_\Lambda,D_\Sigma,D_\Xi)=(-3.08,-3.22,-3.37,-3.45)$
in the convention used here~\cite{Won2022}. In our calculation the
$\Sigma$ instead has the smallest $|D(0)|$. The holographic
momentum radii~\cite{Liu2026} have the opposite
$\Lambda$--$\Sigma$ ordering. Some light-cone sum-rule central
fits for the quark EMT give $r_A^\Sigma>r_A^\Lambda$ and
$r_D^\Sigma>r_D^N$~\cite{Ozdem2020,Azizi2020}.


\section{Flavor breaking and the stress radii}
\label{sec:flavor-breaking}
\label{sec:su3-radius-trajectory}

To follow the flavor dependence of the gravitational form factors, we vary
the strange current mass along
\begin{equation}
 m_s(\lambda)=m_l+\lambda(m_s^{\rm phys}-m_l),\qquad
 0\leq\lambda\leq1.
 \label{eq:su3-current-path}
\end{equation}
We use the same interaction couplings and proper-time cutoffs
along this path, with the gravitational vertices defined in
section~\ref{sec:sources}. At each sampled value we
solve the gap and diquark pole equations and calculate the residues.
We then determine the baryon masses, amplitudes
and normalizations and evaluate the same total EMT.

At $\lambda=0$ the octet is degenerate. The nucleon contains no strange
line in this truncation, so its form factors stay constant along the
trajectory. At the four nonzero values
$\lambda=0.25,0.50,0.75,1$, the $\Sigma$ has a smaller $r_A$ than
the nucleon. The difference $r_D^\Sigma-r_D^N$ is negative at
$\lambda=0.25$ and positive at the three larger values.

We also calculate the electromagnetic radii with the same diquark
poles as in the main form-factor calculation. With a bare constituent
photon vertex, we include the contact terms, contributions from the
proper-time integration limits and scalar--axial-vector transitions.
The charged-baryon radii are
\begin{equation}
 (r_E^p,r_E^{\Sigma^+},r_E^{\Sigma^-},r_E^{\Xi^-})
 =(0.644,0.643,0.562,0.511)~\mathrm{fm},
\end{equation}
where $r_E^2=-6(\hbar c)^2G_E'(0)/Q_B$ for nonzero charge
$Q_B$. The near equality of the proton and $\Sigma^+$ electric
radii contrasts with their separated momentum and stress radii.
The two currents weight the same states differently, as also seen
in unified meson studies~\cite{Xu2024}. Diquark correlations
likewise affect the flavor dependence of nucleon electromagnetic
form factors~\cite{Cloet2014}.


\section{Stress distributions at finite resolution}
\label{sec:mechanics}

For the spin-averaged Breit-frame matrix element, the spatial
stress tensor is
\begin{equation}
 \frac{\langle T_H^{ij}\rangle_{\rm av}}{2E}
 =\frac{D(Q^2)}{4M_B}(q^iq^j-\delta^{ij}Q^2).
\end{equation}
We use the Breit-frame definition of the stress
distributions~\cite{Lorce2019,Jaffe2021,Panteleeva2021,Panteleeva2023GFF}
and apply a Gaussian smearing factor to each baryon,
\begin{equation}
 \widetilde D_{\Lambda}(r)=\frac{1}{2\pi^2}
 \int_0^{Q_{\max}}\dd Q\,Q^2j_0(Qr)
 e^{-Q^2/\Lambda_{\rm res}^2}D(Q^2),
 \qquad Q_{\max}^2=2~\mathrm{GeV}^2.
 \label{eq:resolved-D}
\end{equation}
The associated pressure and shear force distributions are~\cite{Polyakov2018}
\begin{align}
 p_\Lambda(r)&=\frac{1}{6M_Br^2}\frac{\dd}{\dd r}
 [r^2\widetilde D_\Lambda'(r)],\nonumber\\
 s_\Lambda(r)&=-\frac{r}{4M_B}\frac{\dd}{\dd r}
 [\widetilde D_\Lambda'(r)/r].
 \label{eq:resolved-stress}
\end{align}
We integrate over the calculated momentum range and continue the
first two points linearly to the origin. A prime in
Eq.~\eqref{eq:resolved-stress} denotes a radial derivative.

\begin{figure}[tb]
\centering
\includegraphics[width=\textwidth]{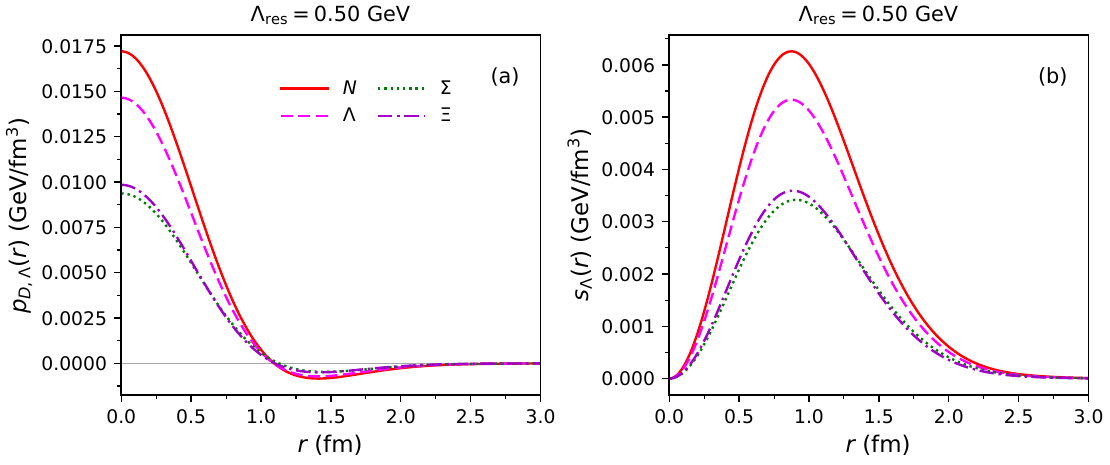}
\caption{Pressure (a) and shear (b) with Gaussian smearing at
$\Lambda_{\rm res}=0.50~\mathrm{GeV}$ and
$Q_{\max}^2=2~\mathrm{GeV}^2$. The baryon conventions follow
Fig.~\ref{fig:gff}.}
\label{fig:mechanics-main}
\end{figure}

All four baryons have positive pressure near the center and negative
pressure at larger radii in Fig.~\ref{fig:mechanics-main}. At the same
resolution, the $\Sigma$ has the weakest pressure and shear peaks.


\section{Summary}
\label{sec:conclusions}

We have calculated the gravitational form factors $A$, $J$ and $D$
of $N$, $\Lambda$, $\Sigma$ and $\Xi$ in the three-flavor NJL
quark--diquark model with proper-time regularization and the
parameters used for the octet spectrum. The total
EMT includes the quark and diquark contributions, the interaction
current and the transitions between scalar and axial-vector diquarks.
For all four baryons, $A$ and $J$ decrease with spacelike momentum
transfer, while $D$ is negative throughout the calculated interval
up to $Q^2=2~\mathrm{GeV}^2$.

The nucleon has the largest $A$ radius, while the $\Sigma$ has the
largest $D$ radius and the smallest $|D(0)|$. The $\Xi$ has the
smallest radius in both cases. Numerically,
$r_D^\Sigma=0.675~\mathrm{fm}$ exceeds the nucleon value
$0.637~\mathrm{fm}$, while $D_\Sigma(0)=-1.091$ is closer to
zero than $D_N(0)=-1.392$. The corresponding $A$ radii are
$0.416$ and $0.423~\mathrm{fm}$. The ordering of the $D$ radii
therefore differs from that of the $A$ radii.

The $\Lambda$ and $\Sigma$ each contain one strange valence quark
and have different gravitational radii. Their spin--flavor wave
functions give different weights to scalar and axial-vector diquarks.
From $\Lambda$ to $\Sigma$, $r_A$ increases from $0.401$ to
$0.416~\mathrm{fm}$ and $r_D$ from $0.624$ to $0.675~\mathrm{fm}$.
We also evaluated the electromagnetic current between these
baryon states. With a bare photon vertex, the proton and $\Sigma^+$
have nearly equal electric radii, $0.644$ and $0.643~\mathrm{fm}$, while their
$A$ and $D$ radii differ. The $\Sigma^-$ and $\Xi^-$ electric radii
decrease to $0.562$ and $0.511~\mathrm{fm}$. The two currents probe
the scalar and axial-vector correlations with different weights.

Varying the strange current-quark mass follows the development of
these splittings from the flavor-symmetric limit. The octet radii
coincide at that limit. At all four nonzero sampled masses,
$r_A^\Sigma<r_A^N$. For $D$, the difference
$r_D^\Sigma-r_D^N$ is negative at $\lambda=0.25$ and positive at
$\lambda=0.50$, $0.75$ and $1$. Flavor breaking thus produces
different splittings in the momentum and stress radii.

For the proton, the model gives smaller $A$ and $D$ radii than
those extracted from the lattice dipole fits. The NJL values are
$0.423$ and $0.637~\mathrm{fm}$,
compared with $0.503$ and $1.145~\mathrm{fm}$ from those fits.
The calculated $D_N(0)=-1.392$ also has a smaller magnitude than
the lattice dipole value $-3.87(97)$.
At the lattice pion mass, the broader forward band from diffusion
reconstruction includes our $D_N(0)$~\cite{AlharazinPanteleeva2026}.

The Breit-frame pressure and shear distributions display the
flavor dependence of the internal stress.
All four baryons have positive pressure near the center and
negative pressure farther out. The $\Sigma$ has the weakest
pressure and shear peaks at the common spatial resolution used here.

More broadly, the octet comparison connects the mechanical structure
of hadrons to flavor symmetry breaking and the correlations that bind
their quark constituents. In the present model, the different flavor
patterns of the momentum and stress radii indicate that these
observables resolve distinct aspects of baryon structure. Extending
the calculation to include long-distance meson dynamics and testing
the dependence on the transverse gravitational current are natural
next steps. Lattice determinations of hyperon GFFs at matched quark
masses would provide a direct test of the predicted octet pattern.
Together, such studies would help determine how quark correlations
and chiral dynamics shape the internal stress of baryons in QCD.


\acknowledgments

This work was supported in part by the National Natural Science Foundation
of China (NSFC) under Grant No.~12603046; the Postdoctoral Fellowship
Program and China Postdoctoral Science Foundation under Grant
Nos.~GZB20260770 and 2026T190847; the JSPS Research Fellowship for Young
Scientists (No.~JP26KJ1338); and the Fundamental Research Funds for the
Central Universities (No.~2026CXGR050).




\end{document}